# Efficient Spectral Broadening in the 100-W Average Power Regime Using Gas-Filled Kagome HC-PCF and Pulse Compression

Florian Emaury,[1,*] Clara J. Saraceno,[1,2] Benoit Debord,[3] Debashri Ghosh,[3] Andreas Diebold,[1] Frederic Gerome,[3] Thomas Südmeyer,[2] Fetah Benabid,[3] and Ursula Keller[1]

[1]*Department of Physics, Institute for Quantum Electronics, ETH Zurich, Wolfgang-Pauli-Str. 16, 8093 Zurich, Switzerland*
[2]*Laboratoire Temps-Fréquence, Université de Neuchâtel, 2000 Neuchâtel, Switzerland*
[3]*GPPMM group, XLIM Research institute, CNRS UMR 7252, University of Limoges, 87060 Limoges, France*
*\*Corresponding author: emaury@phys.ethz.ch*

We present nonlinear pulse compression of a high-power SESAM-modelocked thin-disk laser (TDL) using an Ar-filled hypocycloid-core Kagome Hollow-Core Photonic Crystal Fiber (HC-PCF). The output of the modelocked Yb:YAG TDL with 127 W average power, a pulse repetition rate of 7 MHz, and a pulse duration of 740 fs was spectrally broadened 16-fold while propagating in a Kagome HC-PCF containing 13 bar of static Argon gas. Subsequent compression tests performed using 8.4% of the full available power resulted in a pulse duration as short as 88 fs using the spectrally broadened output from the fiber. Compressing the full transmitted power through the fiber (118 W) could lead to a compressed output of >100 W of average power and >100 MW of peak power with an average power compression efficiency of 88%. This simple laser system with only one ultrafast laser oscillator and a simple single-pass fiber pulse compressor, generating both high peak power >100 MW and sub-100-fs pulses at megahertz repetition rate, is very interesting for many applications such as high harmonic generation and attosecond science with improved signal-to-noise performance.

The development of ultrafast laser systems based on the thin-disk technology has shown impressive progress in the past decade. In particular, semiconductor saturable absorber mirror (SESAM) modelocked thin-disk lasers (TDLs) achieve the highest average power and pulse energy of any ultrafast oscillator. They currently reach pulse energies and average power levels comparable with amplifier systems operating at megahertz repetition rate [1-3], directly from the output of a table-top oscillator. The combination of high average power (up to 275 W [4]) and high pulse energy (up to 80 µJ [5]) makes them very promising for numerous applications. For industrial applications many cold ablation processes benefit from the high peak power and the high repetition rate at few-picosecond pulse durations for an increased throughput [6]. For applications in strong laser field physics or attosecond science, high repetition rates allow for better signal-to-noise ratios or a reduction of space-charge effects [7, 8]. Although significant progress has been achieved in reducing the pulse duration available from these sources [9], the pulse length available directly out of these state-of-the-art oscillators with tens of megawatts of peak power still remains too long (> 500 fs) for the above-mentioned scientific applications. These applications require significant pulse shortening, e.g. for efficient high harmonic generation (HHG).

Efficient compression of these sources that operate with hundreds of watts of average power and high peak powers has remained a challenge for a long time. Large mode area (LMA) silica photonic crystal fibers (PCFs) have shown their capabilities for large compression factors [10] and high average power up to 250 W [11]. However, the input peak power is limited to ≈ 4-6 MW in this configuration due to the self-focusing threshold of silica. This limit is an order of magnitude lower than the output peak power currently reached with TDLs, thus limiting silica PCFs' potential for compression of these state-of-the-art systems. Hollow-core capillaries can be adapted for peak power levels >100 MW and high average powers (≈ 100 W) but exhibit medium overall compression efficiencies (typically below 50%) [12], in particular for small core areas. Temporal and/or spatial beam division [13, 14] has been studied to overcome some of these effects, showing combined average power of hundreds of watts, however, with the trade-off of an increased complexity. A promising alternative for reducing the pulse duration of state-of-the-art modelocked TDLs into the sub-100-fs range is based on spectral broadening and pulse compression in gas-filled Kagome-type hollow-core photonic crystal fibers (HC-PCFs). The intrinsic properties of their guiding mechanism, commonly referred to as "inhibited coupling" (IC) [15], enable a large transmission bandwidth and excellent power handling. In 2010, these properties have been further improved with the introduction of hypocycloid (i.e. negative curvature) core-contour Kagome HC-PCFs [16, 17]. These fibers consequently improve both the high damage threshold and the ultra-low losses over a wide transmission window. The potential of these fibers for pulse compression of TDLs was previously illustrated [18] in a first proof-of-principle result taking the output of a medium power TDL from 860 fs, 7.3 W and 1.9 µJ down to sub-50 fs at 1.1 µJ and 4.1 W of average-power using a 2.8-m long, 16 bar Xenon-filled Kagome HC-PCF [18]. More recently, the delivery and self-compression of record-high 1-mJ-energy-level femtosecond pulses [19] was demonstrated at 1 W of average power. However, it remained unclear whether efficient pulse compression in the multi-100-watt performance level of current TDLs could be obtained using this approach.

Here, we demonstrate that Kagome HC-PCFs are suitable for femtosecond pulse compression in the 100-watt average power and several tens of megawatts peak power regime. The output of our TDL (127 W, 7 MHz, 18 µJ, 740 fs) was directly launched into a Kagome-type HC-PCF statically filled with 13 bar of Argon. With 93% of transmission through the fiber assembly, 118 W of average power were spectrally broadened with very high efficiency. Furthermore, we show the compressibility of our spectrally-broadened output to 88 fs using a fraction of the output of the fiber (8.4%) and negative dispersion mirrors. Dispersion compensation is a linear process and therefore independent of power as long as all optical components withstand the high power operation. Thus we can extrapolate our result to 112 W of equivalent average compressed power. This would result in a source reaching over 88% of overall compression efficiency and a peak power of 105 MW. This performance will enable HHG at high repetition rates in the near future.

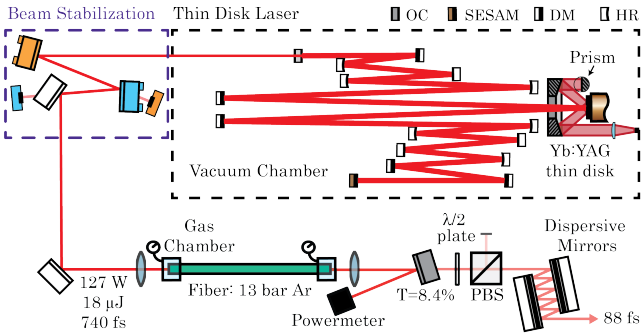

Fig. 1. Experimental setup for the high average power pulse compression. The output of the thin-disk laser (TDL) is spatially stabilized and directly sent into the fiber. The 66 cm-long hypocycloid-core Kagome HC-PCF is held in a gas chamber with 13 bar of Argon. The transmission through the fiber assembly alone (before focusing lens to after collimating lens) reaches up to 118 W of average output power (93% of transmitted power). To show the compressibility of the spectrally-broadened output, 8.4% of the collimated beam is sent through a polarizing beam splitter (PBS) and several bounces on a pair of dispersive mirrors.

The experimental setup is shown in Fig. 1. The Yb:YAG thin-disk oscillator is a modified version of the one presented in [4]. It was built in a vacuum chamber operating at ≈150 mbar of Helium to reduce the nonlinearity of the intracavity environment. Helium was used to ensure good thermal management and reliable daily operation of the oscillator. Dispersive mirrors provided a total negative dispersion per roundtrip of ≈ -10'000 $fs^2$ inside a 21-m long cavity to achieve soliton modelocking. A SESAM was used for starting and stabilizing the soliton modelocking mechanism [20]. In this configuration, we achieve up to 135 W of average output power with 127 W available for compression. With a pulse repetition rate of 7 MHz, this corresponds to a pulse energy of 18 µJ at a pulse duration of 740 fs and 21 MW of peak power. The $sech^2$-shaped pulses were 1.1-times transform-limited with a full-width at half maximum (FWHM) of 1.6 nm (green curve on Fig. 3). In this configuration, the laser operated stably over hours on a daily basis.

We used a single lens with a focal length of 80 mm to focus on the 7-cell 3-ring hypocycloid-core Kagome-type HC-PCF with a spot diameter of ≈ 30 µm (equivalent numerical aperture (NA) < 0.025). A commercially available stabilization kit (Fig. 1) was used to avoid damage of the fiber tip due to slow drifts of the oscillator output beam when the setup was used for long periods of time (> 30 min). Fig. 2a shows a picture of a section of the fiber used in this experiment. The fiber's hypocycloid contour has an inner diameter $D_{in}$ of 42 µm and an outer diameter $D_{out}$ of 59 µm. The measured focus spot diameter fits well with the expected mode-field diameter (MFD) in the fiber and the relation MFD = 0.7 x $D_{in}$ between the fiber inner diameter and the MFD.

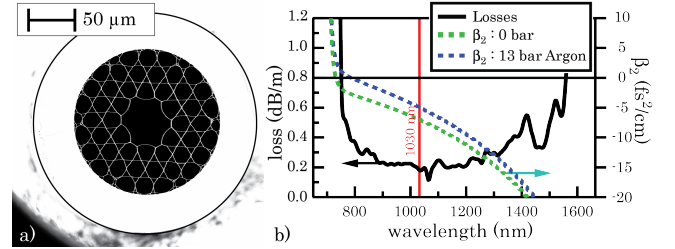

Fig. 2. 7-cell 3-ring hypocycloid core Kagome HC-PCF used in this experiment. a) SEM picture of the fiber showing its core structure ($D_{in}$ = 42 µm, $D_{out}$ = 59 µm). b) Losses and dispersion of the fiber optimized for nonlinear pulse compression. The dispersion of the fiber was calculated using its real effective refractive index and the contribution of 13 bar of Argon.

This fiber has a similar design to the one used for our first compression experiment [18]. It was optimized for operation at our laser wavelength of 1030 nm with a loss figure of 180 dB/km (Fig. 2b). A lower loss figure could be obtained with higher negative curvatures similar to the one used here [21, 22], and further improvement of the already excellent transmission of our setup can be expected using the next generation of fibers (up to 97% through the fiber assembly). Special care was taken to achieve negligible dispersion at the wavelength of 1030 nm (≈-7.0 $fs^2$/cm) in the case of an empty fiber. The 66 cm-long piece of fiber was then placed in a sealed holder where different gases could be used with a pressure up to 15 bar. We chose Argon because of its high ionization threshold and for its relatively high nonlinear refractive index ($n_{2,Ar}$ = 1.4x$10^{-23}$ $m^2$/(W·bar) [23]) to achieve sufficient spectral broadening. More importantly, operating at relatively high gas pressure allowed us to red-shift the zero dispersion wavelength (ZDW) and to use a lower dispersion value (Fig. 2b). Even though the fiber was operated in the anomalous dispersion regime, spectral broadening via self-phase modulation (SPM) still occurred without the onset of solitonic dynamics [19].

Tailoring the fiber to a normal dispersion in the future will enable a better post-compression management.

The 66-cm-long fiber statically filled with 13 bar of Argon showed 93% of transmission over the full range of input average powers, reaching up to 118 W of output average power. This is to our knowledge the first time that more than 100 W of average power were transmitted through such a fiber. Furthermore, this leads to a unique combination of high average power (>100 W), high peak power (>20 MW) and high intensity (> 5 TW/cm$^2$) inside a fiber length of 66 cm, opening new possibilities in nonlinear optics at high average power.

During our experiment, the transmitted output power from the fiber was stable within 1% for more than 30 minutes without the need to realign the fiber coupling, nor to water-cool the fiber holder. No damage or degradation of the fiber was observed after several days of use. This proves the outstanding average and peak power handling of this type of fiber, despite the very thin web cladding structure (Fig. 2a) where the thickness of the inward silica struts was 140 nm and for the outward ones was 350 nm. Further improvement of the fiber mounting and thermal management of the holder should allow for transmission and compression of even much higher average power through such a fiber.

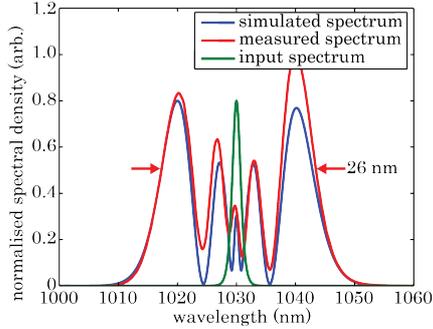

Fig. 3. Output spectrum of the fiber filled with 13 bar of Argon. The input of nearly transform-limited pulses with a FWHM of 1.6 nm (740 fs) (green curve) are spectrally broadened to 26 nm of FWHM (red curve) mostly based on self-phase modulation (SPM). The simulated spectrum (blue curve) shows good agreement with the measured spectrum.

The spectrum of our laser (1.6 nm FWHM) is broadened by SPM more than 16-times to 26 nm FWHM. Fig. 3 shows the broadened spectrum of the fiber output at a launched input power of 127 W. Using a split-step Fourier method, we simulated the broadening occurring in the fiber taking only into account SPM and dispersion [18]. For this simulation we assumed a Gaussian mode-field diameter (MFD) of 30 μm. The inherent dispersion of the fiber (Fig. 2b) was computed using the modal solver of the commercial software Comsol Multiphysics 3.5 based on the finite-element method with an optimized anisotropic perfectly matched layer (PML) [24]. The dispersion of the 13 bar argon-filled fiber is then determined following the typical method to calculate dispersion in capillary fibers [25]. At 13 bar of Argon, the ZDW can be estimated to be around 786 nm, and the small group velocity dispersion (GVD) at 1030 nm is -4.9 fs$^2$/cm which corresponds to 0.87 ps/(nm.km). In our split-step Fourier simulation, we used the SPM factor γ as a free parameter. Good agreement with the measurements was obtained (Fig. 3) by lowering the initial γ value by 40%. This is most likely due to small discrepancies in the exact value of the MFD and the value of the nonlinear refractive index of Argon itself, since this value was measured at 800 nm with 120 fs pulses [23].

Linear pulse compression was demonstrated using only 8.4% of the collimated output beam of the fiber. We then used a polarizing beam splitter (transmission >95%) to ensure a clean linear output polarization (Fig. 1). For compression of the spectrally broadened pulses, we used dispersive mirrors centered at 1030 nm with a group delay dispersion (GDD) of -550 fs$^2$ per reflection. The best compression was achieved with 20 reflections with a total GDD of -11 000 fs$^2$ and a calculated transmission of 99.7% over 20 bounces. For safety and convenience reasons, the compression experiments were realized using 8.4% of the power of the output beam (Fig. 1). These components have previously shown that they can sustain >200 W of average power and such a compression scheme is a linear process independent of the average power. Thus, compressing the full power of the output of the fiber will be straightforward. Taking into account the transmission through the fiber assembly (93%) and the transmission of our compression stage (polarization cleaning = 95% and transmission through the dispersive mirrors = 99.7%), we are confident to reach a highly efficient overall compression of 88% in terms of average power with potentially more than 100 W of compressed average power.

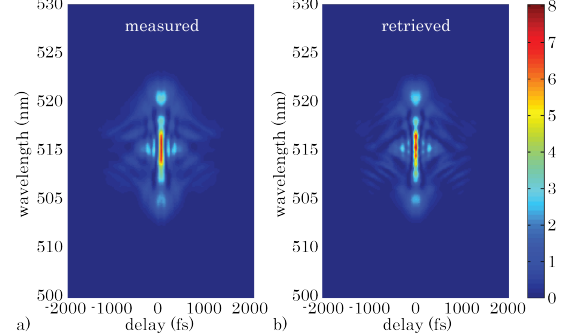

Fig. 4. SHG-FROG measured (a) and retrieved trace (b) in linear scale. The FROG grid used was 512x512 with a temporal resolution of 7.5 fs and a span of 4 ps. The FROG error was below 4.3.10$^{-3}$.

At the optimum dispersion configuration, a SHG FROG trace was taken using a home-built device (Fig. 4). The intensity profile of the pulse in the time domain reaches a FWHM pulse duration of 88 fs with 59% of energy in the main pulse (Fig. 5). The transform-limited pulses from the measured broadened spectrum support 84 fs with 80% of energy in the main peak. Therefore, the compression quality can be further improved using compression mirrors with better dispersion profiles, which were not available at the time of the experiment. Nevertheless, in our current setup, we reach a compression factor of 8.4 with respect to pulse duration in this linear compression regime.

We have shown for the first time that IC guiding Kagome-type HC-PCFs are excellent candidates for femtosecond pulse compression in the 100-watt average power regime. We spectrally broadened the output of a SESAM-modelocked TDL delivering 740-fs pulses at an average power of 127 W, reaching 93% of transmission through the fiber assembly and an output power of 118 W supporting 84 fs pulses. We have demonstrated pulse compression with pulse durations as short as 88 fs using 8.4% of the available power. Applying this compression scheme on the full output of the fiber will allow us in a straightforward setup to reach an overall average power efficiency of 88%. The resulting source with a peak power in excess of 100 MW and a pulse duration of sub-100 fs will be very well-suited for the generation of high-order harmonics at high repetition rates in the near future. Further optimization of the dispersive mirrors as well as the gas pressure in the fiber in our current layout will allow us to increase the energy in the main compressed peak and to reach even shorter pulse durations. Furthermore, an improved thermal management of the fiber will enable even higher average power of several hundreds of watts.

We would like to acknowledge financial support by the Swiss National Science Foundation (SNSF) for this work.

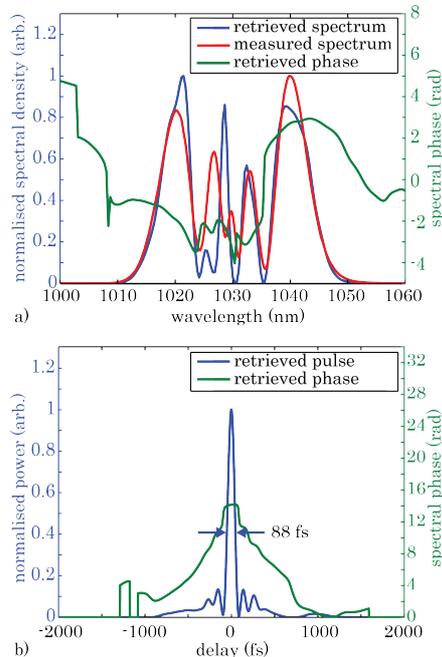

Fig. 5. Retrieved characteristics of the linear pulse compression using 8.4% of the available output power of the fiber. a) FROG-retrieved spectrum and measured spectrum are in good agreement. b) The compressed pulses reached a FWHM pulse duration of 88 fs with 59% of energy in the main peak.